\documentclass[12pt]{article}

\usepackage{epsfig}                                       
               
\let\vf=\vfil \let\cl=\centerline              

\let\a=\alpha  \let\g=\gamma                  
\let\e=\varepsilon  \let\h=\eta                
   \let\m=\mu               
\let\n=\nu  \let\p=\pi \let\r=\rho \let\s=\sigma                
 \let\t=\perp

 \let\G=\Gamma                                
                                              
\def\0{\over } \def\1{\vec }     \def\2{{1\over2}} \def\4{{1\over4}}     
\def\5{\bar }  \def\6{\partial } \def\7#1{{#1}\llap{/}}                  
\def\8#1{{\textstyle{#1}}}       \def\9#1{{\bf {#1}}}                    
 \def\llp{\hbox to 0pt{\hss /\hskip1.5pt}} 
\def\llo{\hbox to 0.2pt{\hss /}} \def\llq{\hbox to 0pt{\hss /\hskip0.5pt}}   
\def\so{\supset\hbox to 0pt{\hss $\displaystyle -$\hskip1pt}}

\def\<{\langle } \def\>{\rangle }

\def\bea{\begin{eqnarray}} \def\eea{\end{eqnarray}}                       
\def\beann{\begin{eqnarray*}} \def\eeann{\end{eqnarray*}}                
\def\beq{\begin{equation}} \def\eeq{\end{equation}}

\renewcommand{\p}{{\scriptscriptstyle \|}}

\def\a{\alpha}
\def\b{\beta}

\def\e{\epsilon}
\def\f{\phi}
\def\g{\gamma}
\def\h{\eta}

\def\l{\lambda}
\def\m{\mu}
\def\n{\nu}

\def\p{\pi}

\def\r{\rho}
\def\s{\sigma}
\def\t{\tau}

\def\G{\Gamma}

\def\vf{\varphi}

\def\cl{{\cal L}}

\def\co{{\cal O}}

\def\bo{{\raise.15ex\hbox{\large$\Box$}}}               
\def\pr{\prod}                                          
\def\face{{\raise.2ex\hbox{$\displaystyle \bigodot$}\mskip-2.2mu \llap {$\ddot
        \smile$}}}                                      
\def\dg{\dagger}                                     

\def\VEV#1{\left\langle #1\right\rangle}        
\def\leftrightarrowfill{$\mathsurround=0pt \mathord\leftarrow \mkern-6mu
        \cleaders\hbox{$\mkern-2mu \mathord- \mkern-2mu$}\hfill
        \mkern-6mu \mathord\rightarrow$}       
\def\dvec#1{\vbox{\ialign{##\crcr
        \leftrightarrowfill\crcr\noalign{\kern-1pt\nointerlineskip}
        $\hfil\displaystyle{#1}\hfil$\crcr}}}           

\def\beq{\begin{equation}}
\def\eeq{\end{equation}}

\def\beqx{\begin{displaymath}}
\def\eeqx{\end{displaymath}}

\def\beqa{\begin{eqnarray}}
\def\eeqa{\end{eqnarray}}
\def\NO{\nonumber}

\def\pl#1#2#3{Phys.~Lett.~{\bf B {#1}} (19{#2}) #3}
\def\np#1#2#3{Nucl.~Phys.~{\bf B {#1}} (19{#2}) #3}
\def\prl#1#2#3{Phys.~Rev.~Lett.~{\bf #1} (19{#2}) #3}
\def\pr#1#2#3{Phys.~Rev.~{\bf D {#1}} (19{#2}) #3}

\def\mpl#1#2#3{Mod.~Phys.~Lett.~{\bf A {#1}} (19{#2}) #3}
\def\nc#1#2#3{Nuovo Cim.~{\bf {#1}} (19{#2}) #3}

\date{}
\title{
{\large\rm DESY 96-158}\hfill{\large\tt ISSN 0418-9833}\\
{\large\rm August 1996}\hfill\vspace*{3cm}\\
Baryon Asymmetry and Neutrino Mixing}
\author{W. Buchm\"uller and M. Pl\"umacher\\
\vspace{3.0\baselineskip}                                               
{\normalsize\it Deutsches Elektronen-Synchrotron DESY, 22603 Hamburg, Germany}
\vspace*{2cm}\\                     
}                                                                          
\input psfig                                                                   
\begin{document}                                                  

\setlength{\baselineskip}{18pt}                                     
\maketitle  
\begin{abstract}
\noindent
In theories where $B-L$ is a spontaneously broken local symmetry, the
cosmological baryon asymmetry can be generated by the out-of-equilibrium
decay of heavy Majorana neutrinos. We study this mechanism assuming a
similar pattern of mixings and masses for leptons and quarks, as suggested
by SO(10) unification. This implies that $B-L$ is broken at the unification 
scale $\Lambda_{\mbox{\scriptsize GUT}}\sim 10^{16}$ GeV,
if $m_{\n_\m} \sim 3\cdot 10^{-3}$eV as preferred by the MSW explanation of
the solar neutrino deficit. The observed value of the baryon asymmetry,
$n_B/s \sim 10^{-10}$, is then obtained without any fine tuning of parameters.
\end{abstract} 
\thispagestyle{empty}
\newpage 

                                            
The standard model of electroweak interactions and its unified extensions 
contain the ingredients which are necessary to explain the observed 
cosmological baryon asymmetry \cite{kt}. However, despite much effort
during almost twenty years, the origin of the baryon asymmetry has not
yet been unequivocally identified. Unified theories, with or without
supersymmetry, offer several interesting scenarios, but it proved
difficult to satisfy all constraints imposed by the dynamics of the
cosmological expansion, which appears to require an inflationary period.
So far no `standard model' of baryogenesis has emerged.

At temperatures above the critical temperature of the electroweak phase
transition baryon ($B$) and lepton ($L$) number violating processes are in
thermal equilibrium \cite{sphal2}. This observation is of crucial 
importance for the theory of baryogenesis. In principle, it opens the
possibility to generate the baryon asymmetry at the electroweak
phase transition \cite{ckn}. However, as a result of detailed studies
of the thermodynamics of this transition in recent years, this now appears 
unlikely, at least within the standard model \cite{jan}.

At high temperatures, where baryon and lepton number violating processes 
are in thermal equilibrium, a baryon asymmetry can be generated from
a lepton asymmetry. This was suggested by Fukugita and Yanagida \cite{fy2}.
The primordial lepton asymmetry is generated by the out-of-equilibrium
decay of heavy Majorana neutrinos in the standard manner. This
mechanism has subsequently been studied by several authors 
\cite{luty,etc,pluemi}, and it has been shown that the observed baryon 
asymmetry,
\beq
     Y_B={n_B\over s}=(0.6-1)\cdot10^{-10}\, ,
\eeq
can be obtained for a wide range of parameters. 

In the high temperature phase of the standard model the asymmetries of
baryon number $B$ and of $B-L$ are proportional in thermal equilibrium 
\cite{sphal},
\beq
     Y_B=\left({8N_f+4N_H\over22N_f+13N_H}\right)Y_{B-L}\, .
\eeq
Here $N_f$ is the number of quark-lepton families and $N_H$ is the
number of Higgs doublets. In the standard model, as well as its
unified extension based on the group SU(5), $B-L$ is conserved. Hence,
no asymmetry in $B-L$ can be generated, and $Y_B$ vanishes.
Furthermore, as mentioned above, baryogenesis at the electroweak phase
transition appears unlikely. As a consequence, the non-vanishing of
the baryon asymmetry is a strong argument for lepton number violation.
This is naturally realized by adding right-handed Majorana neutrinos
to the standard model. This extension of the standard model can be
embedded into grand unified theories with gauge groups containing
SO(10) \cite{fri}. Heavy right-handed Majorana neutrinos can also
explain the smallness of the light neutrino masses via the see-saw
mechanism \cite{seesaw}.

In unified theories with right-handed neutrinos $B-L$ is spontaneously
broken. In this paper we study the implications of baryogenesis on the
scale of $B-L$ breaking and on CP violating leptonic interactions.
Adding right-handed neutrinos to the standard model introduces many
new parameters. We shall restrict this freedom by assuming a similar
pattern of mixings and masses for leptons and quarks, which is natural
in SO(10) unification.

Masses and couplings of charged leptons and neutrinos are given by the
lagrangian
\beq
  \cl_Y = -\overline{l_L}\,\tilde{\f}\,g_l\,e_R
          -\overline{l_L}\,\f\,g_{\n}\,\n_R
          -{1\over2}\,\overline{\n^C_R}\,M\,\n_R
          +\mbox{ h.c.}\;,
\eeq
where $l_L=\left(\n_L,e_L\right)$ is the left-handed lepton doublet
and $\f=(\vf^0,\vf^{-})$ is the standard model Higgs doublet. The
vacuum expectation value of the Higgs field $\VEV{\f}=v\ne0$ generates
Dirac masses $m_l$ and $m_D$ for charged leptons and neutrinos,
\beq
     m_l=g_lv \quad\mbox{and}\quad
     m_D=g_{\n}v\;,
\eeq
which are assumed to be much smaller than the Majorana masses $M$.
Therefore, we have light and heavy neutrinos
\beq
     \n\simeq K^{\dg}\n_L+\n_L^C K\quad,\quad
     N\simeq\n_R+\n_R^C\, ,
\eeq
with masses
\beq
     m_{\n}\simeq- K^{\dg}m_D{1\over M}m_D^T K^*\,
     \quad,\quad  m_N\simeq M\, ,
     \label{seesaw}
  \eeq
as mass eigenstates. Here $K$ is a unitary matrix which relates weak and
mass eigenstates. Since the heavy neutrinos $N_i$ are Majorana fermions, 
they violate lepton number if they decay to lepton and Higgs scalar. 
In the rest system the decay width of $N_i$ reads at tree level,
\beq
    \G_{Di}:=\G_{rs}\left(N^i\to\f^{\dg}+l\right)+
    \G_{rs}\left(N^i\to\f+\overline{l}\right)=
    {M_i\over8\p}{(m_D^{\dg}m_D)_{ii}\over v^2}\;.
    \label{decay}
\eeq
Interference between tree level and one-loop amplitudes yields the
$CP$ asymmetry \cite{pluemi}
\beqa
  &&\e_i={1\over8\pi v^2\left(m_D^{\dag}m_D\right)_{ii}}\sum\limits_j
  \mbox{Im}\left[\left(m_D^{\dag}m_D\right)_{ij}^2\right]\,f
  \left({M_j^2\over M_i^2}\right)\label{cpasymm}\\[1ex]
  &&\quad\mbox{with}\quad f(x)=\sqrt{x}\left[1-(1+x)\ln\left({1+x\over x}
  \right)\right]\;.\NO
\eeqa
In a quantitative analysis of this mechanism one has to take into
account several other processes as well, especially the lepton number violating
scatterings mediated by a massive neutrino $N_i$. In the following we shall 
take all three heavy  neutrino families into account as intermediate states, 
but we shall only calculate the asymmetry generated by the lightest of the
right-handed neutrinos, since the asymmetries generated by the heavier
neutrinos are washed out.

{\noindent\it Neutrino masses and mixings}

In this paper we make the ansatz of a similar pattern of mixings and
mass ratios for leptons and quarks, which is natural in SO(10)
unification.  Such an ansatz is most transparent in a basis where all
mass matrices are maximally diagonal. In addition to real mass
eigenvalues two mixing matrices appear.  We can always choose a basis
for the lepton fields such that the mass matrices $m_l$ for the
charged leptons and $M$ for the heavy Majorana neutrinos $N_i$ are
diagonal with real and positive eigenvalues,
  \beq
  m_l=\left(\begin{array}{ccc}m_e&0&0\\0&m_{\m}&0\\0&0&m_{\t}
  \end{array}\right)\qquad
  M=\left(\begin{array}{ccc}M_1&0&0\\0&M_2&0\\0&0&M_3
  \end{array}\right)\;.
  \eeq
In this basis $m_D$ is a general complex matrix, which can 
be diagonalized by a biunitary transformation. Therefore, we can
write $m_D$ in the form
  \beq
  m_D=V\,\left(\begin{array}{ccc}
  m_1&0&0\\0&m_2&0\\0&0&m_3\end{array}\right)\,U^{\dag}\;,
  \eeq
where $V$ and $U$ are unitary matrices and the $m_i$ are real and
positive. In the absense of a Majorana mass term $V$ and $U$ would 
correspond to Kobayashi-Maskawa type mixing matrices of left- and 
right-handed charged currents, respectively.

According to eqs.~(\ref{decay}) and (\ref{cpasymm}) the $CP$ asymmetry
is determined by the mixings and phases present in the product
$m_D^{\dag}m_D$, where the matrix $V$ drops out.  Therefore, to
leading order, the mixings and phases which are responsible for
baryogenesis are entirely determined by the matrix $U$.
Correspondingly, the mixing matrix $K$ in the leptonic charged
current, which determines $CP$ violation and mixings of the light
leptons, depends on mass ratios and mixing angles and phases of $U$
and $V$.  Hence, there is no direct connection between the $CP$
violation and generation mixing at high and low energies.

We now concentrate on the mixing matrix $U$. One can factor out five phases,
which yields
  \beq
    U=\mbox{e}^{i\g}\,\mbox{e}^{i\l_3\a}\,\mbox{e}^{i\l_8\b}\,U_1\,
    \mbox{e}^{i\l_3\s}\,\mbox{e}^{i\l_8\t}\;,
  \eeq
where the $\l_i$ are the Gell-Mann matrices. The remaining matrix
$U_1$ depends on three mixing angles and one phase, like the 
Kobayashi-Maskawa matrix for quarks. In analogy to the quark mixing
matrix we choose the Wolfenstein parametrization \cite{wolfenstein} as
ansatz for $U_1$,
  \beq\label{mm}
    U_1=\left(\begin{array}{ccc}
    1-{\l^2\over2}  &      \l        & A\l^3(\r-i\h)\\[1ex]
        -\l         & 1-{\l^2\over2} & A\l^2 \\[1ex]
    A\l^3(1-\r-i\h) &    -A\l^2      &  1
    \end{array}\right)\;,
  \eeq
where $A$ and $|\r+i\h|$ are of order one, while the mixing
parameter $\l$ is assumed to be small. For the masses $m_i$ and
$M_i$ we assume a hierarchy like for up-type quarks,
  \beqa
  m_1=b\l^4m_3&\quad m_2=c\l^2m_3&\quad b,c=\co(1)\\[1ex]
  M_1=B\l^4M_3&\quad M_2=C\l^2M_3&\quad B,C=\co(1)\;.\label{Mmass}
  \eeqa
For the eigenvalues $m_i$ of the Dirac mass matrix this choice is
motivated by SO(10) unification. The masses $M_i$ cannot be
degenerate, because in this case there exists a basis for $\n_R$
such that $U = 1$, which implies that no baryon asymmetry is
generated. For simplicity we therefore assume that the masses $M_i$
scale like the Dirac neutrino masses.

The light neutrino masses are given by the seesaw formula
(\ref{seesaw}). The matrix $K$, which diagonalises the neutrino mass
matrix, can be evaluated in powers of $\l$. A straightforward calculation
gives the following masses for the light neutrino mass eigenstates
  \beqa
     m_{\n_e}&=&{b^2\over\left|C+\mbox{e}^{4i\a}\;B\right|}\;\l^4
             \;m_{\n_{\t}}+\co\left(\l^6\right)\label{mne}\\[1ex]
     m_{\n_{\m}}&=&{c^2\left|C+\mbox{e}^{4i\a}\;B\right|\over BC}
             \;\l^2\;m_{\n_{\t}}+\co\left(\l^4\right)\label{mnm}\\[1ex]
     m_{\n_{\t}}&=&{m_3^2\over M_3}+\co\left(\l^4\right)\;.\label{mnt}
  \eeqa

The $CP$-asymmetry in the decay of the lightest right-handed
neutrino $N_1$ is easily obtained from eqs.~(\ref{cpasymm}) and
(\ref{mm})-(\ref{Mmass}),  
  \beq
    \e_1=-\;{1\over16\p}\;{B\;A^2\over c^2+A^2\;|\r+i\h|^2}\;\l^4\;
    {m_3^2\over v^2}\;\mbox{Im}\left[(\r-i\h)^2
    \mbox{e}^{i2(\a+\sqrt{3}\b)}\right]
    \;+\;\co\left(\l^6\right)\;.
  \eeq
This yields for the magnitude of the $CP$ asymmetry,
  \beq\label{cpa}
    |\e_1| \leq {1\over16\p}\;{B\;A^2\;|\r+i\h|^2\over c^2+A^2\;|\r+i\h|^2}\;
    \l^4\;{m_3^2\over v^2}\;+\;\co\left(\l^6\right)\;.
  \eeq
How close the value of $|\e_1|$ is to this upper bound depends on the phases
$\a$, $\b$ and $\arg{(\r+i\h)}$. 
Since $\e_1\propto m_3^2/v^2$, we can already conclude that a large value
of the Yukawa coupling $m_3/v$ will be preferred by this mechanism of 
baryogenesis. This holds irrespective of the neutrino mixings.

\pagebreak
{\noindent\it Numerical results \label{results}}

To obtain a numerical value for the produced baryon asymmetry, we have to
specify the free parameters in our ansatz (\ref{mm})-(\ref{Mmass}). 
In the following we will always use as a constraint the value for 
the $\n_{\m}$-mass which is preferred by the 
MSW explanation \cite{msw} of the solar neutrino deficit (cf.~\cite{kir}),
  \beq
    m_{\n_{\m}}\simeq 3\cdot10^{-3}\;\mbox{eV}\;. \label{msw}
  \eeq
A generic choice for the free parameters is to take all $\co(1)$
parameters equal to one and to fix $\l$ to a value which is
of the same order as the $\l$ parameter of the quark mixing matrix,
  \beq
    A=B=C=b=c=|\r+i\h|\simeq 1\; ,\qquad \l\simeq 0.1\;. \label{p1}
  \eeq
{}From eqs.~(\ref{mne})-(\ref{mnt}), (\ref{msw}) and (\ref{p1}) one
now obtains,
\beq
  m_{\n_e}\simeq 8\cdot10^{-6}\;\mbox{eV}\; , \quad
  m_{\n_{\t}}\simeq 0.15\;\mbox{eV}\; .\label{m1}
\eeq
Finally, a second mass scale has to be specified. In unified theories based 
on SO$(10)$ the Dirac neutrino mass $m_3$ is naturally equal to the 
top-quark mass. Hence, we choose
  \beq\label{3t}
     m_3=m_t\simeq 174\;\mbox{GeV}\;.
  \eeq
This determines the masses of the heavy Majorana neutrinos $N_i$,
  \beq
     M_3 \simeq 2\cdot10^{14}\;\mbox{GeV}\; ,\label{M3}
\eeq
and, consequently, $M_1\simeq 2\cdot10^{10}\;\mbox{GeV}$ and 
$M_2\simeq 2\cdot10^{12}\;\mbox{GeV}$. {}From eq.~(\ref{cpa}) one obtains
the $CP$ asymmetry $|\e_1| \simeq 10^{-6}$, where we have assumed maximal 
phases. The solution of the set of Boltzmann equations discribed in 
\cite{pluemi} now yields the $B-L$ asymmetry (see fig.~1a),
  \beq
     Y_{B-L} \simeq 3\cdot10^{-10}\; ,
  \eeq
which is indeed the correct order of magnitude. The precise value depends
on unknown phases.

  \begin{figure}
     \begin{minipage}[t]{8cm}
     \mbox{ }\hfill\hspace{1cm}(a)\hfill\mbox{ }
     \epsfig{file=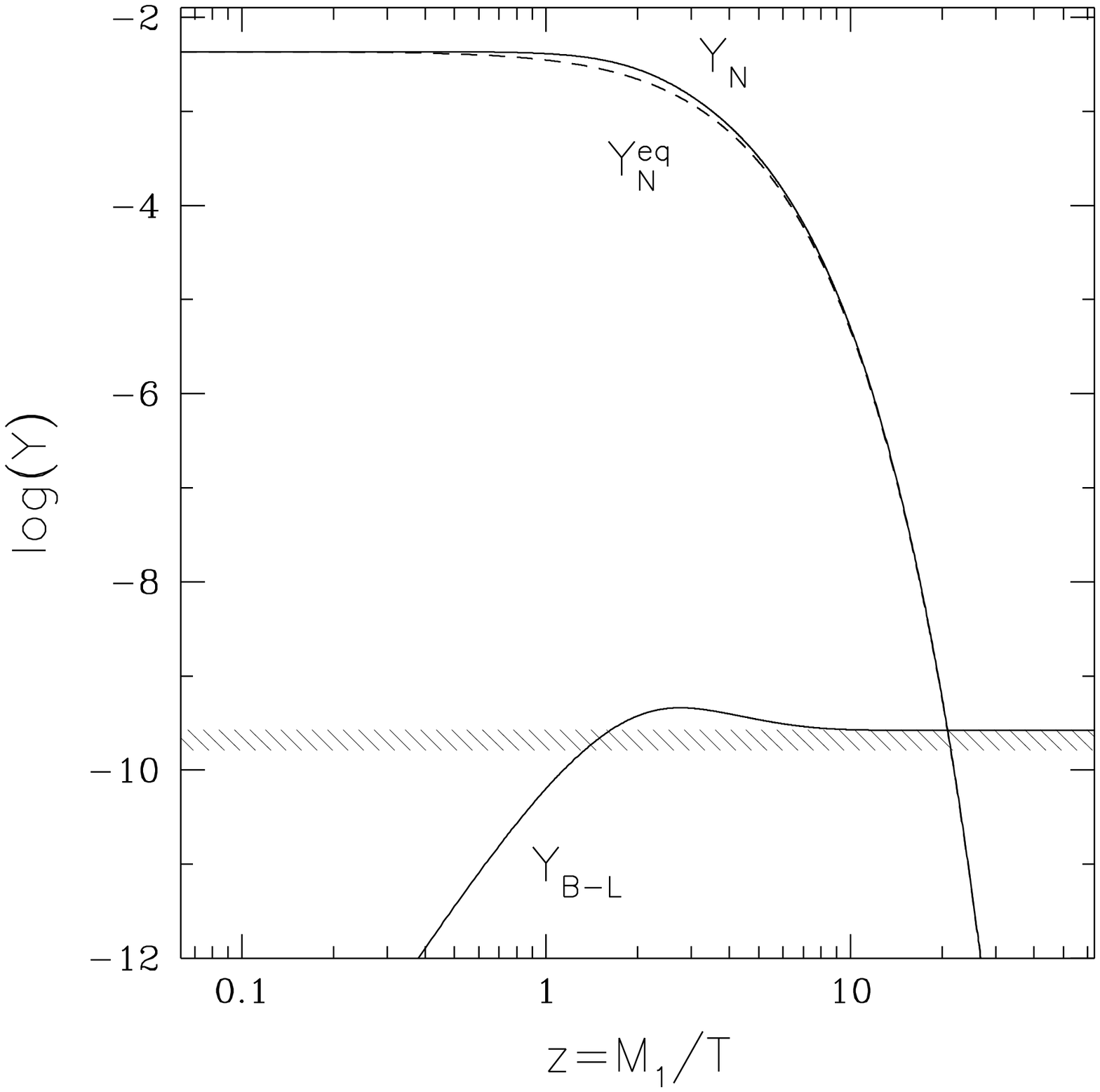,width=8cm}
     \end{minipage}
     \hspace{\fill}
     \begin{minipage}[t]{8cm}
     \mbox{ }\hfill\hspace{1cm}(b)\hfill\mbox{ }
     \epsfig{file=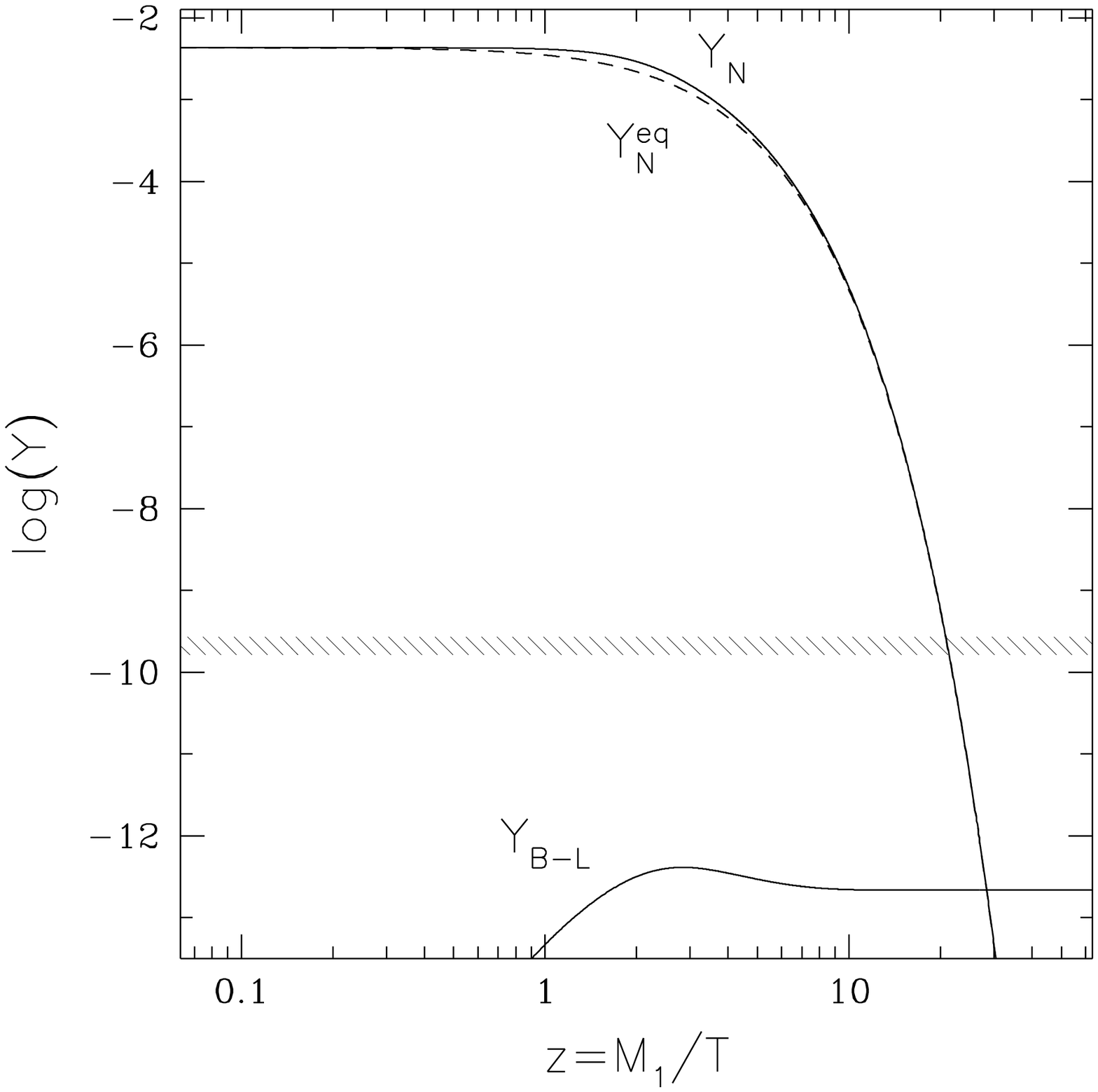,width=8cm}
     \end{minipage}  
     \caption{\it Time evolution of the neutrino number density and
     the $B-L$ asymmetry for $\l=0.1$ and for $m_3=m_t$ (a)
     or $m_3=m_b$ (b). The equilibrium distribution for
     $N_1$ is represented by a dashed line, while the hatched area
     shows the measured value for the asymmetry.}
  \end{figure}

The large mass $M_3$ of the heavy Majorana neutrino $N_3$ (cf.~(\ref{M3})), 
suggests that $B-L$ is already broken
at the unification scale $\Lambda_{\mbox{\scriptsize GUT}} \sim 10^{16}$
GeV, without any intermediate scale of symmetry breaking. The large
value of $M_3$ is a consequence of our choice (\ref{3t}), $m_3 \simeq
m_t$. To test the sensitivity of our result for 
$Y_{B-L}$ on this assumption, consider the alternative choice,
\beq
m_3 = m_b \simeq 4.5\; \mbox{GeV}\; ,
\eeq
with all other parameters remaining unchanged. In this case one obtains
$M_3=10^{11}$ GeV and $|\e_1| = 5\cdot10^{-10}$ for the mass of $N_3$
and the $CP$ asymmetry, respectively. Since the maximal $B-L$ asymmetry is
$-\e_1/g*$, where $g*$ is the number of relativistic degrees
of freedom (cf.~\cite{kt}), it is clear that the generated asymmetry
will be too small. The solutions of the Boltzmann equations are shown
in fig.~1b. The generated asymmetry,
  \beq
     Y_{B-L} \simeq 2\cdot10^{-13}\;,
  \eeq
is too small by more than two orders of magnitude. We can conclude that 
high values for both masses $m_3$ and $M_3$ are preferred, which is
natural in SO(10) unification.

\begin{figure}[t]
    \mbox{ }\hfill
     \epsfig{file=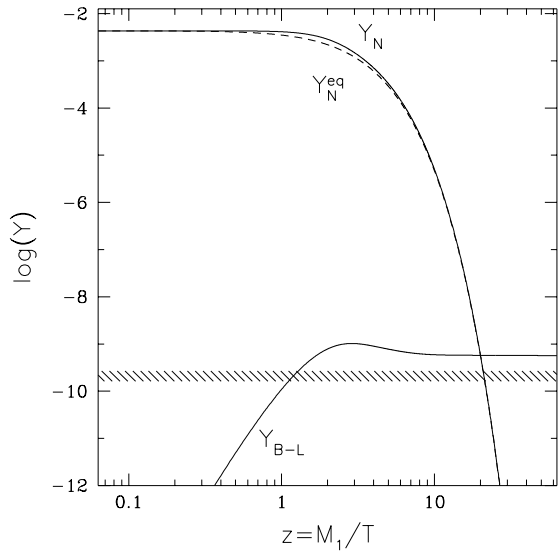,width=8cm}
    \hfill\mbox{ }
   \caption{\it Solutions of the Boltzmann equations for
   $m_{\n_{\t}}=5\;$eV and $m_3=m_t$. \label{fig2}}
\end{figure}

Models for dark matter involving massive neutrinos favour a  
$\t$-neutrino mass \cite{raf},
\beq\label{lnt}
m_{\n_{\t}} \simeq 5\; \mbox{eV}\; ,
\eeq
which is significantly larger than the value given in (\ref{m1}).
The large value (\ref{lnt}) for the $\t$-neutrino mass does not
correspond to the simplest choice of parameters within our ansatz.
However, it can be accomodated for the following set of parameters:
$b=|\r+i\h|\simeq 1$, $A=c\simeq 1/3$, $B=C\simeq 3$, $\l\simeq 0.09$,
$m_3\simeq m_t$. In this case one obtains
$M_3 \simeq 6\cdot 10^{12}$ GeV, $m_{\n_e} \simeq 6\cdot10^{-5}$ eV, 
$|\e_1| \simeq 2\cdot10^{-6}$. Integration of the Boltzmann equations 
yields the $B-L$ asymmetry (see fig~2), 
  \beq
     Y_{B-L} \simeq 6\cdot10^{-10}\; ,
  \eeq
where again maximal phases have been assumed.

We conclude that in unified theories based on the group SO(10), the
out-of-equilibrium decay of heavy Majorana neutrinos naturally explains
the cosmological baryon asymmetry. Assuming a similar pattern of mixings
and masses for leptons and quarks 
the observed value of
the baryon asymmetry is obtained without any fine tuning of parameters.
To leading order in gauge and Yukawa couplings the $CP$ violating
phases, which are relevant at high and low energies, decouple. $B-L$
is broken at the unification scale.

Without an intermediate scale of symmetry breaking, the unification
of gauge couplings appears to require low-energy supersymmetry. This
provides further sources for generating a $B-L$ asymmetry \cite{cam}, 
whose size depends on additional assumptions. In this case, especially
constraints on the reheating temperature \cite{kt} and the 
possible role of preheating \cite{linde} require further studies.

%

\pagebreak

\end{document}